\documentclass[aps,reprint,twocolumn,floatfix,superscriptaddress]{revtex4-1}%,twocolumn,draft
\usepackage{amsmath}
\usepackage{amssymb}
\usepackage{amsthm}
\usepackage{color}
\usepackage{graphics}
\usepackage{graphicx}
\usepackage{dcolumn}
\usepackage{braket}

\newtheorem{theorem}{Theorem}

\newtheorem{definition}{Definition}

\begin{document}

%\title{Raman scattering from bound photons in the ultrastrong coupling regime}
\title{Scattering in the ultrastrong regime: nonlinear optics with one photon}
\author{E. Sanchez-Burillo}
\affiliation{Instituto de Ciencia de Materiales de Aragon and Departamento de Fisica de la Materia Condensada, CSIC-Universidad de Zaragoza, E-50012 Zaragoza, Spain}
\author{D. Zueco}
\affiliation{Instituto de Ciencia de Materiales de Aragon and Departamento de Fisica de la Materia Condensada, CSIC-Universidad de Zaragoza, E-50012 Zaragoza, Spain}
\affiliation{Fundacion ARAID, Paseo Maria Agustin 36, E-50004 Zaragoza, Spain}
\author{J. J. Garcia-Ripoll}
\affiliation{Instituto de Fisica Fundamental, IFF-CSIC, Calle Serrano 113b, Madrid E-28006}
\author{L. Martin-Moreno}
\affiliation{Instituto de Ciencia de Materiales de Aragon and Departamento de Fisica de la Materia Condensada, CSIC-Universidad de Zaragoza, E-50012 Zaragoza, Spain}

\begin{abstract}
The scattering of a flying photon by a two-level system ultrastrongly coupled to a one-dimensional 
photonic waveguide is studied numerically.  The photonic medium is modeled as an array of coupled 
cavities and the whole system is analyzed beyond the rotating wave approximation using Matrix Product States. 
It is found that the scattering is strongly influenced by the single- and multi-photon dressed bound 
states present in the system. In the ultrastrong coupling regime a new channel for inelastic scattering appears, where an incident photon deposits energy into the qubit, exciting a photon-bound state, and escaping with a lower frequency. This single-photon nonlinear frequency conversion process can reach up to 50\% efficiency. Other remarkable features in the scattering induced by counter-rotating terms are a blueshift of the reflection resonance and a Fano resonance due to long-lived excited states
\end{abstract}

%\pacs{42.25.Bs, 41.20.Jb, 42.79.Ag, 78.66.Bz}

\maketitle

{\it Introduction.-} 
As light-matter interaction controls an immense variety of physical processes, its modification usually 
leads to new phenomena. One strategy to increase this interaction is to confine the electromagnetic 
field in waveguides 
%(which can be made in photonics crystals, dielectrics, plasmonic materials, 
%superconducting circuits, a chain of coupled cavities, etc),  
and make it interact  with few level systems.
%(as, for instance, quantum dots, molecules, NV-centers or flux qubits). 
It is possible nowadays 
to reach in this way the situation where the coherent light-matter coupling predominates over decoherence 
processes (the so-called strong-coupling regime), and to generate, manipulate and 
storage a single (or a few) photon. The ability of performing tasks with just one photon 
has already been demonstrated\cite{Astafiev2010, Hoi2011}, 
opening the path for proposals such as 
optical transistors\cite{Fan2003,Lukin2007,Nori2008},  one-photon lasers\cite{Fan2012}, qubit-mediated entanglement\cite{Tudela2011} or efficient photo-detectors\cite{Romero2009a}. 

All these results have been analyzed within the rotating-wave-approximation (RWA) for the photon-dipole interaction\cite{Cohen-Tannoudji1992}. The RWA only considers the processes where light and matter exchange excitations, which is valid when the couplings are much smaller than the typical photon and qubit energies. For sufficiently strong couplings processes involving spontaneous creation and annihilation of pairs of excitations are relevant and the RWA  picture breaks down\cite{DeLiberato2008}. This regime of ultrastrong coupling, opens the door to new physics\cite{LeHur12,Romero12}, which is within reach for many different experimental implementations\cite{Niemczyk2010}.

From the theoretical viewpoint, within the RWA the scattering of multiphoton wavepackets by qubits is a complex problem\cite{Fan2007,Fan2010,Fan2011,Baranguer2014, Longo2009, Longo2010}, but the one-photon scattering is trivial. 
Beyond the RWA computing the scattering of even one flying photon is difficult as subspaces with
different photon numbers mix in the dynamics. This converts the problem into a many-body one for which 
only partial solutions exist for models that consider linear (unbounded) dispersion relations and, typically, either in the perturbative regime ($g/\omega < 0.2$) or in the localization phase, ($g/\omega > 1$)

In this letter we analyze the scattering by one flying waveguide photon by one qubit for 
an ample range of photon-qubit interactions that comprise the strong- and ultra- strong coupling 
regimes, and taking into account effect of non-linearity in the photon dispersion relation. 
For that, we use 
the framework of Matrix Product States\cite{Vidal2003, Vidal2004, Cirac2004} to compute
the many-body dynamics. 
For sufficiently small couplings, we recover the RWA results where the qubit acts as a perfect mirror in resonance. 
However, for stronger couplings, a richer phenomenology is found: 
renormalization of the resonant frequency, appearance of an asymmetric Fano resonance and existence of inelastic Raman processes. 

%%%%%%%%%%%%%%%%%%
%%%%%%%%%%%%%%%%%%
%%%%% Model
%%%%%%%%%%%%%%%%%%
%%%%%%%%%%%%%%%%%%
\begin{figure}[h]
\includegraphics[width=1.0\columnwidth]{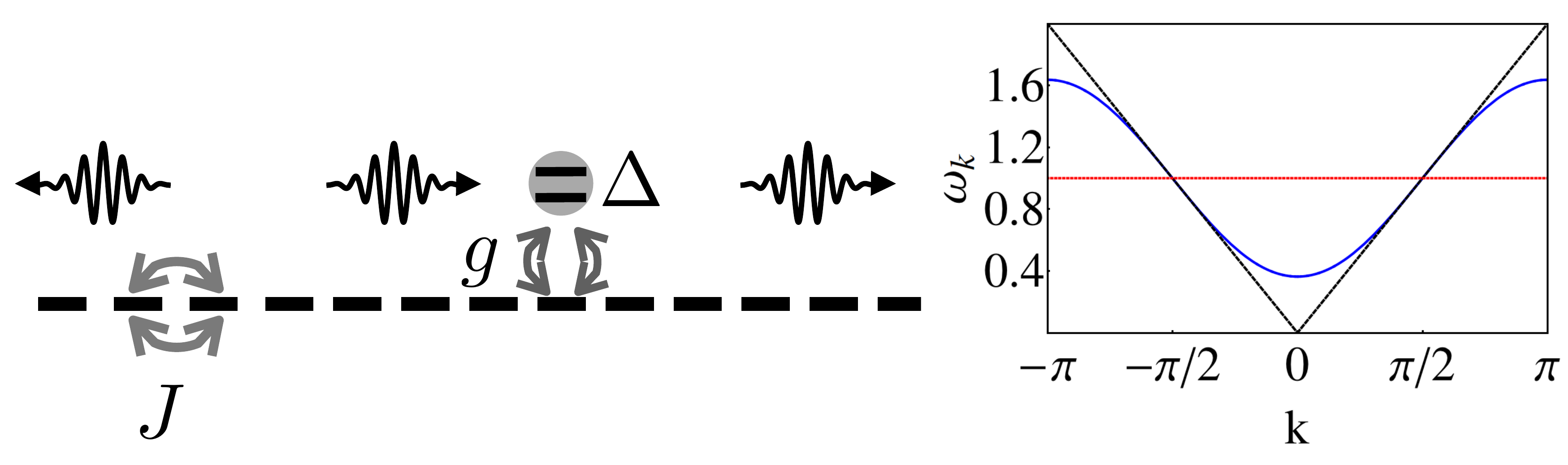}\\
\caption{{\it Left:} Schematics for the considered system. {\it Right:} Dispersion relation of the free photon band considered in this problem. The discontinuous line represents a linear dispersion with the same velocity at the qubit frequency $\Delta=1$}\label{fig:f1}
\end{figure}

{\it Model and methods.-} The photonic medium is represented as a chain of $L$ discrete bosonic sites (which can be considered either as bona-fide coupled cavities or as a discretization of a continuous waveguide) coupled to a qubit living at site $j_0=0$
(see Fig.\ \ref{fig:f1}). 
The Hamiltonian of the combined system is ($\hbar =1$):
%\begin{align}
%\label{eq:H}
%%\nonumber
%H &= \sum_i^L 
%(a^\dagger_i a_i
%+ J (a_{i+1}^\dagger a_i + \mathrm{H.c.}))
% + \Delta \sigma_z  
%%+ g \sigma_x (a_{x_o}^\dagger + a_{x_0} )
%%+ g (\sigma^{+} +\sigma^{-}) X_{i_0}
%+ g \sigma_x X_{i_0}
%\end{align}
%
\begin{equation}
\label{eq:H}
%\nonumber
H = \sum_j 
(a^\dagger_j a_j
+ J (a_{j+1}^\dagger a_j + \mathrm{H.c.}))
 + \Delta \sigma^+\sigma^-  
%+ g \sigma_x (a_{x_o}^\dagger + a_{x_0} )
%+ g (\sigma^{+} +\sigma^{-}) X_{i_0}
+ g \sigma_x X_{j_0}
\end{equation}
where the first two terms represent the photons in the waveguide, the
third one describe the qubit and the fourth term is the interaction
between a dipole transition and the local electric field (characterized by a strength $g$).
In Eq. (\ref{eq:H}), $a^\dagger_{j}$ and $a_{j}$  create and annihilate, respectively, a photon at position $j$,  $X_{j_0} \equiv a_{j_0}^\dagger + a_{j_0}$ and  $\sigma_x$ and $\sigma^\pm$ are Pauli and ladder matrices acting onto the qubit, which has an excitation energy $\Delta$.  
The free-photon dispersion relation depends on both the on-site photon energy (which is taken as the frequency unit) and the hopping parameter $J$:  $\omega_k = 1  + 2 J \cos(k)$. Throughout the paper we take  $J=-1/\pi$ and $\Delta=1$, so that the qubit resonance sits where the photon band is linear. 
Importantly, the finite bandwidth of the dispersion relation implies the existence of bound states localized in the vicinity of the qubit\cite{John90}. As shown below, these states are essential in some scattering properties, so continuum models with unbounded photon dispersion relations may present different physics. Notice though that realistic waveguides always have at least low-frequency cutoffs.
%Using $\sigma_\pm = \sigma_x \pm i \sigma_y$, 

The interaction Hamiltonian can be expressed as the sum of ``rotating wave'' and ``counter-rotating" contributions,  $H_{int}^{RW} = g \, (\sigma^{+}a_{j_0} +  \sigma^{-}a^\dagger_{j_0})$  and $H_{int}^{CR} = g \, (\sigma^{+} a^\dagger_{j_0} + \sigma^{-} a_{j_0}) $, respectively. For $g \ll \Delta$, $H_{int}^{CR}$ can be safely neglected, which greatly simplifies the calculations as $H_{int}^{RW}$ conserves the total number of excitations, $N_{exc}$. For large enough couplings (as a rule of thumb when  $g \geq 0.1 \Delta$) $H_{int}^{CR}$ cannot be neglected and subspaces with different number of excitations are visited during the dynamics. Nevertheless, the full Hamiltonian still has parity 
$\Pi = (-1)^{N_{exc}}$ as a conserved quantity.

As mentioned, the presence of
counter-rotating terms converts the scattering of even a single photon into a many-body
problem. Hence a brute-force computation of the time evolution is prohibitive,
even for small chain lengths.
Our calculations use the representation of 
Matrix Product States (MPS) to describe the wavefunction\cite{JJRipoll2006,Verstraete2008}.
Whenever a many-body state is slightly entangled, as typically occurs for 1D
systems in the low energy sector\cite{Eisert10}, MPS is optimal. The complexity is not exponential
anymore, as it happens for a random state, but it is polynomial with the size of
the system. This allows the study of the low energy physics by means of classical computation. 
Further details on both the method and the tests performed can be found in the Supplementary Material.

The simulation of the scattering process follows the following
steps: {\it (i)} computation of the the ground state (GS), {\it (ii)}
generation of the input state comprising the GS plus one incoming
photon {\it (iii)} time evolution of the wave function and {\it (iv)} analysis of the final wave function.  

\begin{figure}
\includegraphics[width=0.8\columnwidth]{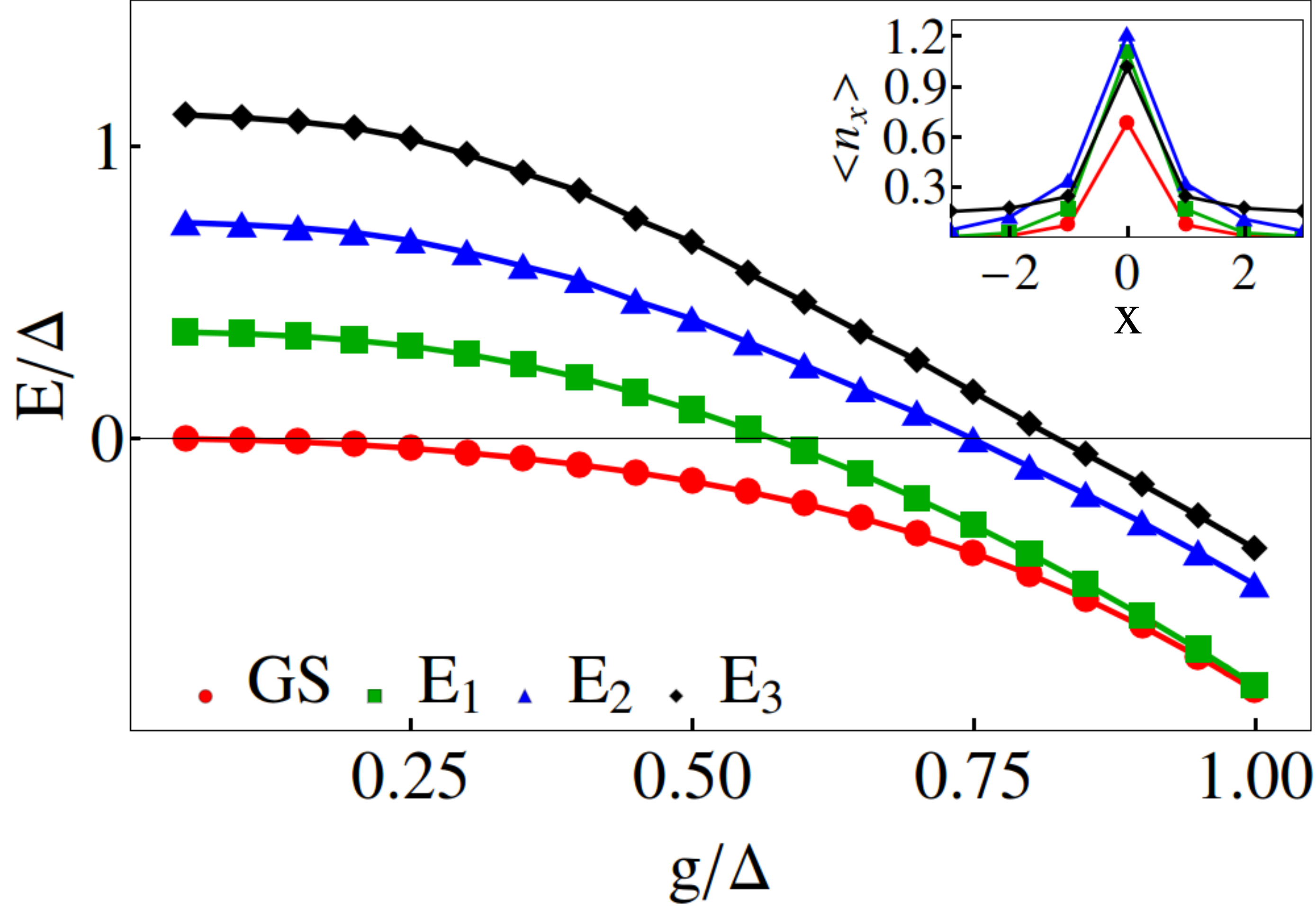}\\
\caption{{\bf Energies of bound states.} Dependence with coupling $g$ of ground state and bound
excited state energies. $GS$ and $E_{2}$ have even parity, while $E_{1}$ and $E_{3}$ have odd parity. The inset shows the spacial profile of the number of excitations in each bound state, for $g=0.8$. }\label{fig:static}
\end{figure}

{\it Ground and excited states.-}
We compute the GS by imaginary time evolution of a seed state. 
Within the RWA the GS is the vacuum (0 photons and 0 qubit
excitations). However, when counter-rotating terms are relevant, the GS is a 
non-trivial ``dressed qubit'', with a
photon cloud bound to the two-level system. Excited bound states can also be computed 
by the same method, by proper 
orthogonalization with lower lying states. 
Figure\ \ref{fig:static} shows the energy of the ground state and the first bound excited states, 
as a function of $g$. Their spatial profile of number of photons in the cloud is rendered in 
the inset to Fig. \ \ref{fig:static}. 

For small $g$, i.e., within the RWA, the index $n$ in $E_{n}$ labels the number of excitations 
in the state (with $GS\equiv E_{0})$. The single-photon bound state $E_1$, already 
predicted in RWA models\cite{John90},  does 
not play a role in the scattering process, as it lies outside the one-photon band. 
On the contrary, the $E_3$ energy lies inside that band, with which it hybridizes. 
Thus, strictly speaking, $E_3$ is a leaky bound mode. 
This complicates the computation of $E_3$ using MPS; 
Fig.\ \ref{fig:static} shows its 
estimated energy, obtained via  numerical
diagonalization of (\ref{eq:H}) for a lattice with $L=7$ sites.  

\begin{figure}[thb!]
\includegraphics[width=.98\columnwidth]{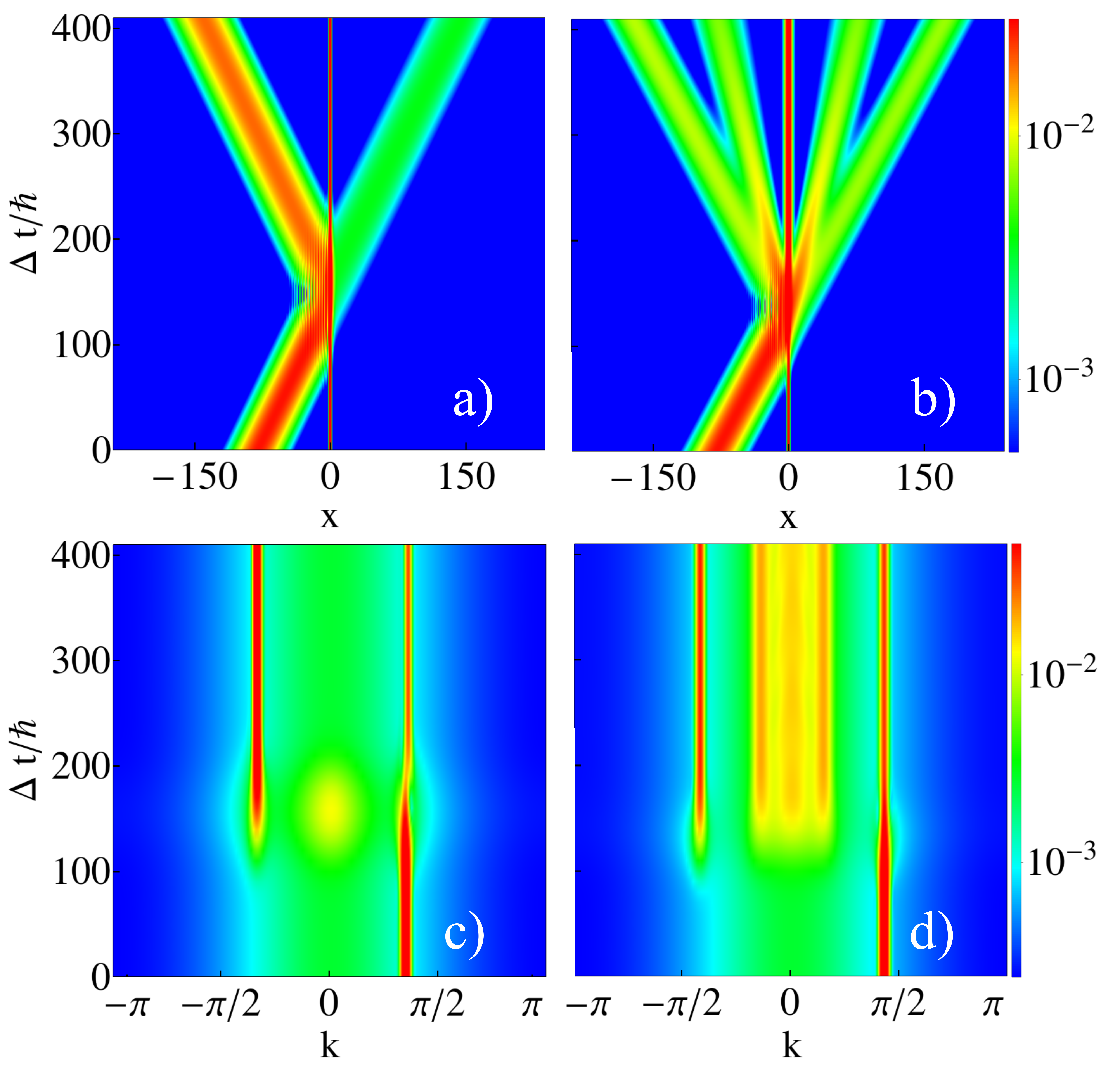}

\caption{{\bf Time evolution.}  
Evolution of $\langle n_x\rangle$ (upper panels) and $\langle n_k \rangle$ (lower panels) 
for $\omega_{in} = 0.70$ (left panels) and $\omega_{in} = 0.85$  (right panels). 
In both cases,  at $t=0$ an initial wave packet is set centered at
$x_0=-80$ and the coupling is $g=0.7$. For $\omega_{in} = 0.70$ the
scattering is elastic, while for $\omega_{in} = 0.85$  there is an
inelastic scattering channeling too.
}\label{fig:frequency_convertion}
\end{figure}

\begin{figure}
\includegraphics[width=.89\columnwidth]{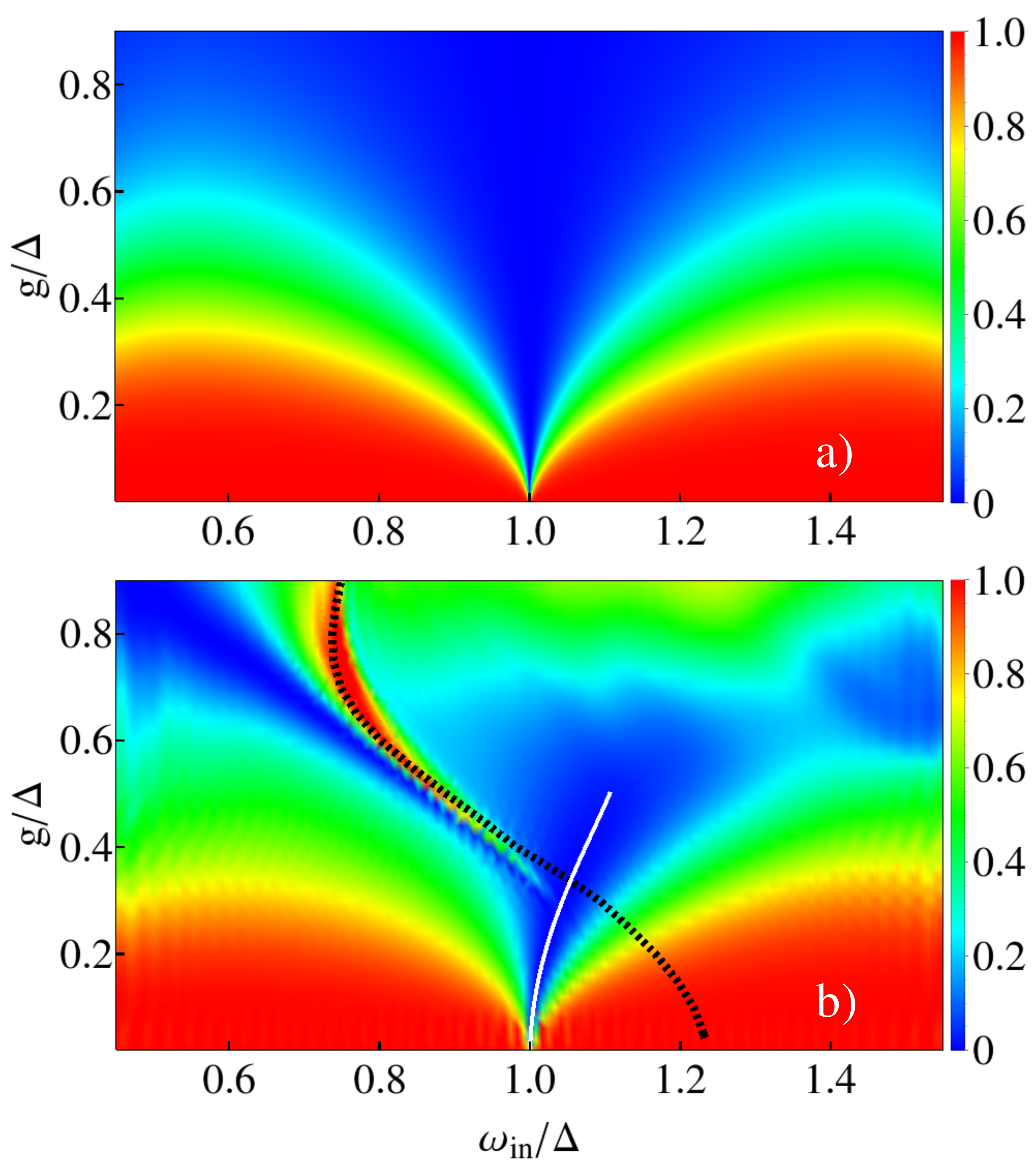}

\caption{{\bf Transmission as a function of both incident photon frequency $\omega_{in}$ and $g$.}  
{\it (a)} Transmittance within the RWA 
{\it (b)}  Elastic Transmittance in the full model. 
The black line marks the estimated frequency for the Fano resonance while the white line gives the estimated spectral position for the transmittance minimum (see text).
%{\it Bottom} Cuts for the Transmission at three selected coupling
%values (the values are given in the plot titles)
}\label{fig:T}
\end{figure}

{\it Scattering simulation.-} 
As input state we create on the GS a one-photon Gaussian wavepacket, centered at $x_0$  with spacial width $\sigma$, moving towards the qubit with average momentum $k_{in}$ (and corresponding frequency $\omega_{in}$), 
\begin{equation}
\ket{\Psi(0)} = a_{\phi}^\dagger \ket{GS} \equiv \sum_{x} \phi_x a_x^\dagger \ket{GS},
\end{equation}
with $ \phi_x \propto \exp[-(x-x_0)^2/2\sigma^2 + i k_{in}]$. The time evolution of this wave gives us $\ket{\Psi(t)}$.

Useful quantities to characterize the scattering are: the average local number of photons 
$\langle  n_x(t) \rangle = \bra{\Psi(t)} a_x^\dagger a_x \ket{\Psi(t)}$, its equivalent in Fourier space $\langle  n_k(t) \rangle$ and the one photon dynamics over the GS $\phi_x(t)=\langle GS| a_x |\Psi(t)\rangle$. From the Fourier transform of the latter we can extract the transmission amplitude as
$t_k = \phi_k(t_{f}) / \phi_k^{free}(t_{f})$,
where $t_{f}$ is a time long enough so that the scattering process has concluded, and $\phi_k^{free}$ is the propagation when the dressed qubit and the incoming photon do not interact.
These quantities suffice for analyzing scattering amplitudes as, in all considered cases, the computed amplitude for generation of more than one propagating photon is negligible.  

Figure \ref{fig:frequency_convertion} shows both $\langle  n_x(t)\rangle$ and 
$\langle  n_k(t)\rangle$ for two representative cases, corresponding to different $k_{in}$, and $g=0.7$. 
For this value of $g$, at which the RWA is not valid, the GS comprises a photon cloud around the qubit,  as seen in both $\langle  n_x(t=0)\rangle$ (at $x \approx 0$) and $\langle  n_k(t=0)\rangle$ 
(which presents a finite value around $k=0$).  As time evolves, we observe the typical scattering evolution. 
After a time span of free propagation ($t \lesssim 100$), an interaction period starts where
both reflected and transmitted photon beams develop. Finally, 
at larger times  ($t\gtrsim 300$), the scattered photon propagates freely. 

There are always reflected and transmitted {\it elastic} beams, which propagate at the same speed 
as the incident one. Remarkably, as shown in the Fig.  \ref{fig:frequency_convertion}(b,d), 
for some parameters there are also inelastic (Raman) processes where both reflected and transmitted wavepackets propagate with a different speed to the incident one (and thus a different frequency). 
Notice also that, 
in this case, after the scattering event the photon cloud around the qubit has changed, broadening in real space (thus narrowing momentum space).

{\it Elastic scattering}
Figure \ref{fig:T} renders the transmission into the {\it elastic} channel, as a 
function of both $\omega_{in}$ and $g$. 
The top panel is obtained within the RWA, while the lower
panel is computed using the full Hamiltonian.

For sufficiently small $g$ ($g\lesssim 0.3$), the elastic transmission spectra is, both within the RWA and for the full model, characterized by a deep transmission minimum, with a spectral width that increases with $g$. The main difference is that, while within the RWA the minimum always occurs at $\omega_{min}=\Delta$, in the full model the transmission minimum blueshifts with $g$. This shift is reminiscent of the frequency renormalization in the
spin-boson model\cite{Guinea1998, Peropadre2013}, which is a continuum model without band edges. However, the renormalization
group flow predicts a redshift of the effective frequency of the qubit. Here, 
the waveguide presents a  natural cutoff at high-$\omega$, which prevents a direct application of 
the renormalization group. 
Nevertheless, in this intermediate regime the counter-rotating terms can be taken into account perturbatively (see Supplementary Material), leading to an analytical condition for the spectral position of the transmission minimum, which is rendered in \ref{fig:T}b (white line).

For larger $g$, ($g>0.3$), an asymmetric  Fano-like resonance develops in the the elastic transmission spectra. 
This feature
combines a deep minimum and a strong transmission maximum, with a line width that increases monotonically with
$g$. Fano resonances are the hallmark of long-lived states entering
the scattering dynamics. 
In this case, its origin can be traced back
to the leaky bound state $E_3$, as shown by the agreement between the frequency at which the resonance occurs and the computed energy difference $E_{3}-E_{GS}$ (black line in figure \ref{fig:static}).  As
commented, within the RWA the state $E_3$  contains three excitations 
and therefore it is not accessible to the propagation of a single photon. Counter-rotating terms mix the one and three excitation sectors, opening the way to the appearance of this novel long-lived transmission resonance.

Notice that for $g\gtrsim 0.7$ a new regime seems to appear where the transmission is largely enhanced for a wide frequency range. This is reminiscent of the decoupling between light and matter predicted when $g\gtrsim 1$ in cavity-QED\cite{deLiberato14}. However, the terms responsible for that decoupling, which involve only photon operators at the qubit position, are not present in our calculation, as they are expected to play a role only for larger $g$'s that those considered here. The analysis of the transmission spectra at such high $g$ values, in the so-called ``deep ultra-strong regime", is an interesting problem that is, however, beyond the aim of this work.

\begin{figure}
\includegraphics[width=.9\columnwidth]{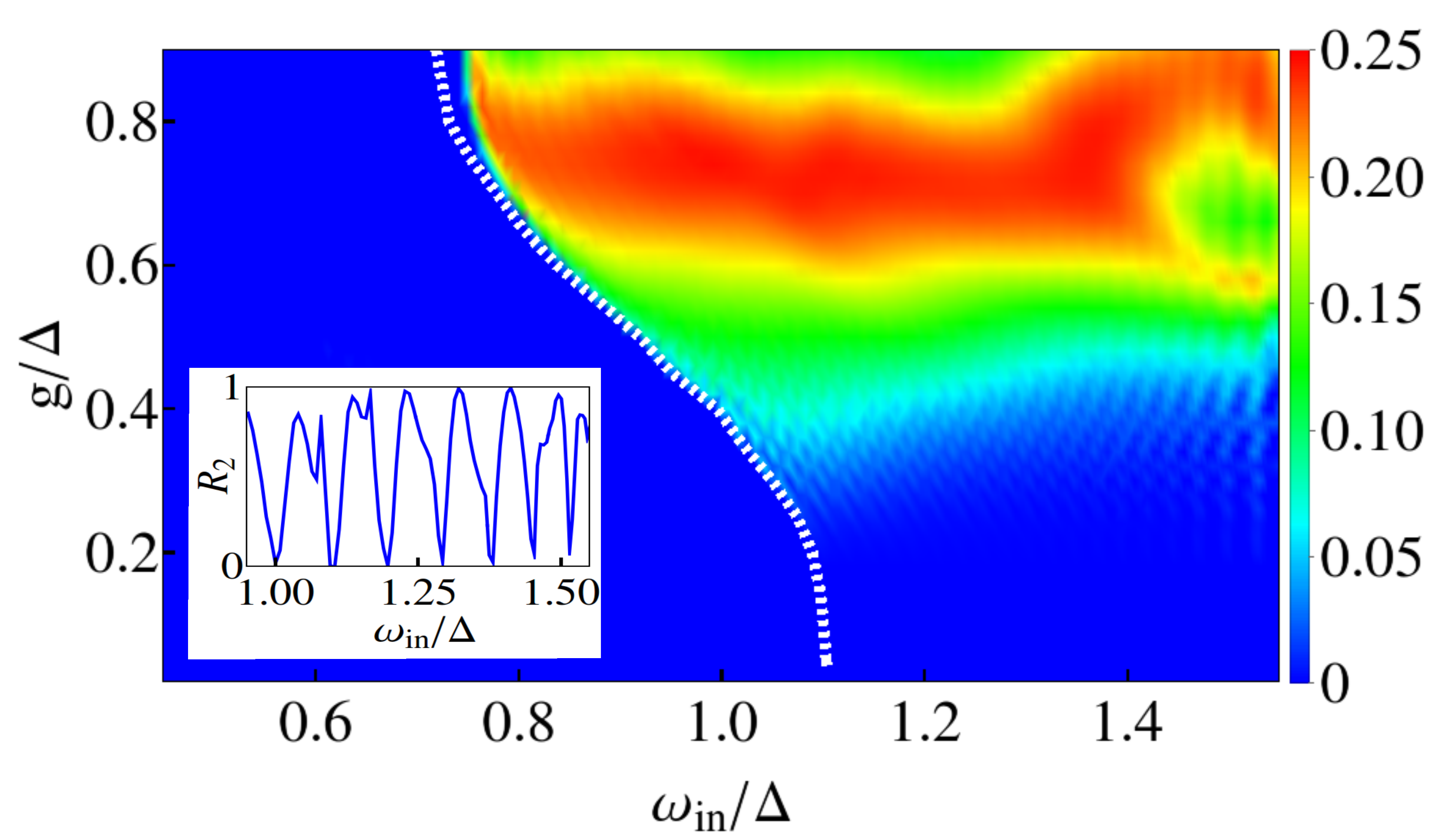}

\caption{{\bf Inelastic transmittance.}  
Transmittance in the full model in the inelastic channel as a function of both incident photon frequency $\omega_{in}$ and $g$.  The white line is
estimated boundary for the region where the photon frequency conversion occurs. The inset presents, for $g=0.8$, the inelastic reflection spectra when the waveguide is terminated at position $\Delta x=20$, showing that 100\% efficient Raman process is possible using one incoming photon. 
%{\it Bottom} Cuts for the Transmission at three selected coupling
%values (the values are given in the plot titles)
}\label{fig:T2}
\end{figure}

{\it Inelastic scattering:  Raman within just one photon.-}
Figure \ref{fig:T2} renders the transmitted flux at frequencies different to the incoming one, as a function of  $\omega_{in}$. The Fourier analysis reveals that the frequency of the output flying photon is linked to  $\omega_{in}$ through 
\begin{equation}
E_{GS} + \omega_{in} = E_{2} + \omega_{\rm out}
\end{equation}
Therefore, this inelastic process corresponds to a Raman scattering\cite{Cohen-Tannoudji1992, Boyd2003}
that leaves the dressed qubit in an excited bound state that, if counter-rotating terms were not present, 
would fully reside in the sector $N_{ext} =2$. Within the RWA this sector is not accessible for 
one photon propagating in the GS, so this Raman process is genuine non-RWA physics. 

As the output flying photon must belong to the one-photon band, the minimum frequency at 
which the Raman process may occur is $\min [\omega_{Raman}] = E_{2}-E_{GS} + 1 - 2 |J|$. 
The dependence with $g$ of this quantity is represented in figure \ref{fig:T2} (white line), 
clearly marking the boundary for existence the inelastic transmission. 

The computed inelastic transmittance never exceeds $0.25$. 
This turns out to be a fundamental upper bound: the maximization of the current in the inelastic
channel, $P_{ine}$, subject to the conditions of current conservation ($ 1-|r|^2-|t|^2=P_{ine}$), 
%where $t$ and $r$ are the 
%transmission and reflection coefficients in the elastic channel) 
and continuity of the photonic wave function ($1+r =t$), readily gives $\max [P_{ine}]=0.5$.
%Energy conservation imposes that the power in the inelastic
%channel, $P_{ine}$, must fulfill $ 1-|r|^2-|t|^2=P_{ine}$, where $t$ and $r$ are the 
%transmission and reflection coefficients in the elastic channel. Maximizing $P_{ine}$, subject to the 
%constrain of continuity of the photonic wave function ($1+r =t$), one gets $\max [P_{ine}]=0.5$. 
As a point-like qubit cannot
differentiate between left and right, $P_{ine}$ is divided equally in both directions.
This argument is analogous to that leading to the maximum possible absorption by point-like 
scatterers.\cite{Romero2009a}
%, Romero2009b}.  
%
Full absorption can be achieved in that case if a mirror is placed in the waveguide 
(the so called ``one-port coherent perfect absorption"\cite{Stone2010}). Exploiting this analogy, 
we have considered the case where the waveguide is terminated at the transmission side of 
the qubit. In this case, ``one-port coherent perfect Raman scattering", implying both photon frequency conversion and 
excitation of the dressed qubit, is possible with unit probability {\it at the one-photon level},
as shown in the inset of Fig. \ref{fig:T2}.
\begin{figure}[thb!]
\includegraphics[width=0.9\columnwidth]{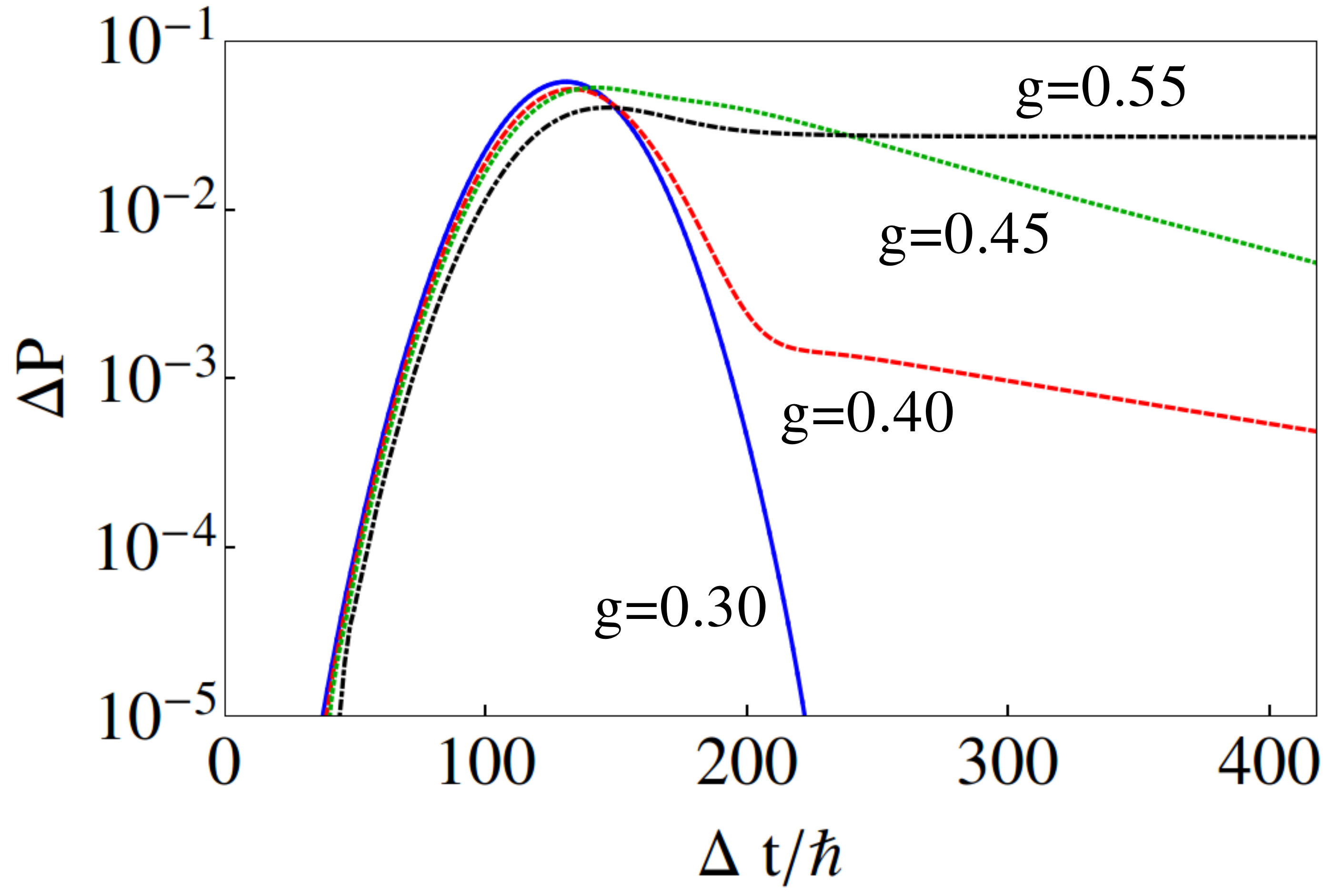}\\
\caption{{\bf Qubit dynamics.}  Time evolution for the population of the qubit 
excited state,
with respect to that in the ground state, for  $\omega_{in} = 0.90$ and several values of the photon-qubit coupling $g$. The wave packet width is $\sigma=20$, for which $\delta \omega_{in}/4J \simeq 0.04 \ll 1$. While for $g=0.30$ (solid)
  the time dynamics corresponds to a fast decay back to the GS, multi-relaxation long-lived process occur for  $g=0.40$ (dashed) and $0.45$ (dotted). At $g=0.55$ (dashed-dotted) Raman scattering is energetically possible and the qubit ends up in an excited stationary state.}\label{fig:qubit}
\end{figure}

It is interesting to analyze whether this Raman process may occur in other systems. It is possible to show that
it cannot occur if the qubit is substituted by a bosonic cavity or resonator, even if the coupling contains 
counter-rotating terms (See Supplementary Material). This negative result can be traced 
back to the linearity of the Heisenberg
equations for the bosonic creation operators. Therefore the system analyzed in this paper represents the
minimal setup for observing inelastic scattering with a single photon.

{\it Time evolution of qubit population.-}
The excitation of dressed-qubit bound states have a strong impact 
on the dynamics of the qubit 
excited state population $P$. Figure \ref{fig:qubit} shows, for several values of the coupling $g$,
the time evolution of 
$\Delta P = P-P_{GS} $ (where $P_{GS}$ is evaluated on the GS)
for an incoming one-photon 
wave packet with a representative $\omega_{in}=0.9$. 
For $g<0.3$, the qubit dynamics is governed by the excitation by the passing wave packet  
and the fast de-excitation of the qubit. For $0.3<g<0.55$, $\Delta P$ shows a slow 
decay characterized by multi-exponential relaxations, associated to the resonant excitation of both $E_3$ and $E_2$
(which is a virtual process in this range). 
For higher $g$, the Raman excitation becomes a real process and $\Delta P$ is finite at long times.

{\it Conclusions.-}
The scattering of a flying photon impinging into a two-level system placed in a waveguide has been
studied for a large range of coupling strengths, including regimes were the rotating-wave approximation 
is no longer valid. For that, we have adapted the technique of Matrix Product States to scattering problems.  
Our results predict a rich phenomenology for the transmission spectra. At sufficiently small photon-qubit couplings
the transmission spectra is dominated by a deep minimum, as found within the RWA. 
But when the coupling is strong enough, we predict new rich phenomenology for the transmission spectra:  
a blueshift in the transmission minima, appearance of  a long-lived Fano resonance 
and highly efficient inelastic processes. All these phenomena are due to the existence of bound multiphoton 
modes which are accessible when the full Hamiltonian is considered.
%
%The single photon-single qubit system turns
%out to be the minimal system for resonant Fano-like transmission and
%Raman scattering with a single photon as an input. 
The explored parameter range is accessible to current experimental state of the art, 
at least using superconducting technology for both qubits and waveguides, thus opening the possibility
to access non-perturbative quantum optics with single or few flying
photons.

{\it Acknowledgements.-}
We acknowledge 
support by the Spanish Ministerio de Economia y Competitividad within projects MAT2011-28581-C02, FIS2012-33022 and No. FIS2011-25167, the Gobierno
de Aragon (FENOL group)
and the European project PROMISCE.

%\bibliographystyle{apsrev4-1}
%\bibliography{scattering_eduardo} 

\appendix

\section{Matrix Product States}

As we indicated in the letter, we solve the problem by using the MPS technique. Let us justify why we can do it.

Unlike in \cite{Peropadre2013}, our bandwidth-limited photonic medium can be treated in the RWA and its ground state is the vacuum both in frequency and position space. Moreover, even if we go beyond RWA in the qubit-resonator coupling ($g/\Delta\gtrsim 0.1$), it is true that the ground state is not the vacuum anymore, as we show in the paper, but it will follow the area law \cite{Eisert10}, so it will be slightly entangled. As we are studying the dynamics of a photon flying over the ground state, the state will have a small amount of entanglement.

The important consequence of the previous discussion is that we may use the variational ansatz of Matrix Product States \cite{JJRipoll2006,Verstraete2008} to describe the discrete wavefunction, since it is valid for 1D systems when the entanglement is small enough. This ansatz has the form
\begin{equation}
\ket{\psi} = \sum_{s_i \in \{1,d_i\}} \mathrm{tr}\left[
\prod A_i^{s_i}\right] \ket{s_1,s_2,\ldots,s_L}.
\label{eq:mps}
\end{equation}
It is constructed from $L$ sets of complex matrices $A_i^{s_i} \in M[\mathbb{C}^{D}]$, where each set is labeled by the quantum state $s_i$ of the corresponding site. The local Hilbert space dimention $d_i$ is infinity, since we are dealing with bosonic sites. However, during the dynamics, processes that create multiple photons are still highly off-resonance. Then, we can truncate the bosonic space and consider states with $0$ to
$n_{max}$ photons per cavity. So, the composite Hilbert space is $\mathcal{H}=\bigotimes_i \mathbb{C}^{d_i}$, where the dimension is $d_i=n_{max}+1$ for the empty resonators and $d_{i_0}=2(n_{max}+1)$ for the cavity with the qubit. We thus expect the composite wavefunction of the photon-qubit system to consist of a superposition with a small number of photons

The total number of variational parameters $(L-1)D^2(n_{max}+1) + 2D^2(n_{max}+1)$ depends on the size of the matrices, $D$. The key point is that, for describing a general state, $D$ increases exponentially with $L$, whereas its dependence is polynomial if the entanglement is small enough, in such a way that the number of parameters increases polynomially with $L$ for this class of states.

Our work with MPS relies on four different algorithms. The most basic one is to create trivial, product states of known shape, such as a vacuum state with a deexcited qubit $\ket{\psi}=\ket{\downarrow}\ket{vac}$. These states can be reproduced using matrices of bond dimension $D=1$, so each matrix is just a coefficient $A_i^{s_i}=\delta_{s_i1}$. The second algorithm is to compute expectation values from MPS. This amounts to a contraction of tensors that can be performed efficiently \cite{JJRipoll2006}, and allows us to compute single-site operators $\langle a^\dagger_i a_i\rangle$, $\langle \sigma_z\rangle$, or correlators, $\langle a_i^\dagger a_j\rangle$. The third operation that we need to perform is to apply operators on to the state, $O\ket{\psi}$, such as introducing or removing excitations $a_i^\dagger\ket{\psi}$. We do this in an efficient fashion by interpreting the operator $O$ as a Matrix Product Operator (MPO) \cite{Verstraete2010}. A MPO is a matrix product representation of an operator:

\begin{equation}
O = \sum_{s_{i}^{},s_{i}^\prime \in \{1,d_i\}} \mathrm{tr}\left[\prod B_i^{s_i^{},s_i^\prime}\right]\ket{s_1^{},s_2^{},\ldots,s_L^{}}\bra{s_1^\prime,s_2^\prime,\ldots,s_L^\prime}
\end{equation}

So, now we have $L$ sets of complex matrices $B_i^{s_i^{},s_i^\prime} \in M[\mathbb{C}^{D_O}]$, where each set is labeled by two indices $s_i^{},s_i^\prime$ of the corresponding site.

We just need to apply sums of one-body operators

\begin{equation}
O = a_\phi^\dagger = \sum_n \phi_n a_n^\dagger.
\end{equation}

In such a case, an efficient representation of the MPO is obtained with $D_O=2$

\begin{equation}
B_i^{s_i^{},s_i^\prime}=\left(\begin{array}{c c}
\delta_{s_i^{},s_i^\prime} & 0\\
\phi_i(a_i^\dagger)_{s_i^{},s_i^\prime} & \delta_{s_i^{},s_i^\prime}
\end{array}\right)\qquad i=2,3,\dots,L-1,
\end{equation}
whereas $B_1^{s_1^{},s_1^\prime}=(\phi_1(a_1^\dagger)_{s_1^{},s_1^\prime},\delta_{s_1^{},s_1^\prime})$ and $B_L^{s_L^{},s_L^\prime}=(\delta_{s_L^{},s_L^\prime},\phi_L(a_L^\dagger)_{s_L^{},s_L^\prime})^T$, with $(a_i^\dagger)_{s_i^{},s_i^\prime}=:\bra{s_i^{}}a_i^\dagger\ket{s_i^\prime}$.

Finally, with this tool in our box, we can also approximate time evolution, repeatedly contracting the state with an MPO approximation of the unitary operator $\exp(-iH\Delta t)$ for short times, and truncating it to an ansatz with a fixed $D$. Since our problem does not contain long-range interactions and since the state is well approximated by MPS, it is sufficient to rely on a third-order Suzuki-Trotter formula \cite{Suzuki1991}.
In the same way as we can consider time evolution, we can take imaginary time to obtain the ground state and excited states, that is solving the equation $i\tfrac{d}{dt}P\ket{\psi}=PHP\ket{\psi}$ for finite time-steps, while constantly renormalizing the state. Here, $P$ is either the identity (for the ground state) or a projector that either selects a well defined quantum number (parity $\Pi$) or projects out already computed states. In either case, provided a suitable initial state, the algorithm converges to the lowest-energy state of $PHP$ in the subspace selected by $P$. Note that, while the excited states are useful in order to interpret the results, the ground state is totally necessary to study the dynamics, since our initial state is just a photon flying over the ground state.

\section{Checking convergence of the algorithm}

In this section, we check that our results converge in the variational parameters: the bond dimension $D$ and the cutoff in the number of excitations per site $n_{max}$.

\begin{figure}[h]
\includegraphics[width=1\columnwidth]{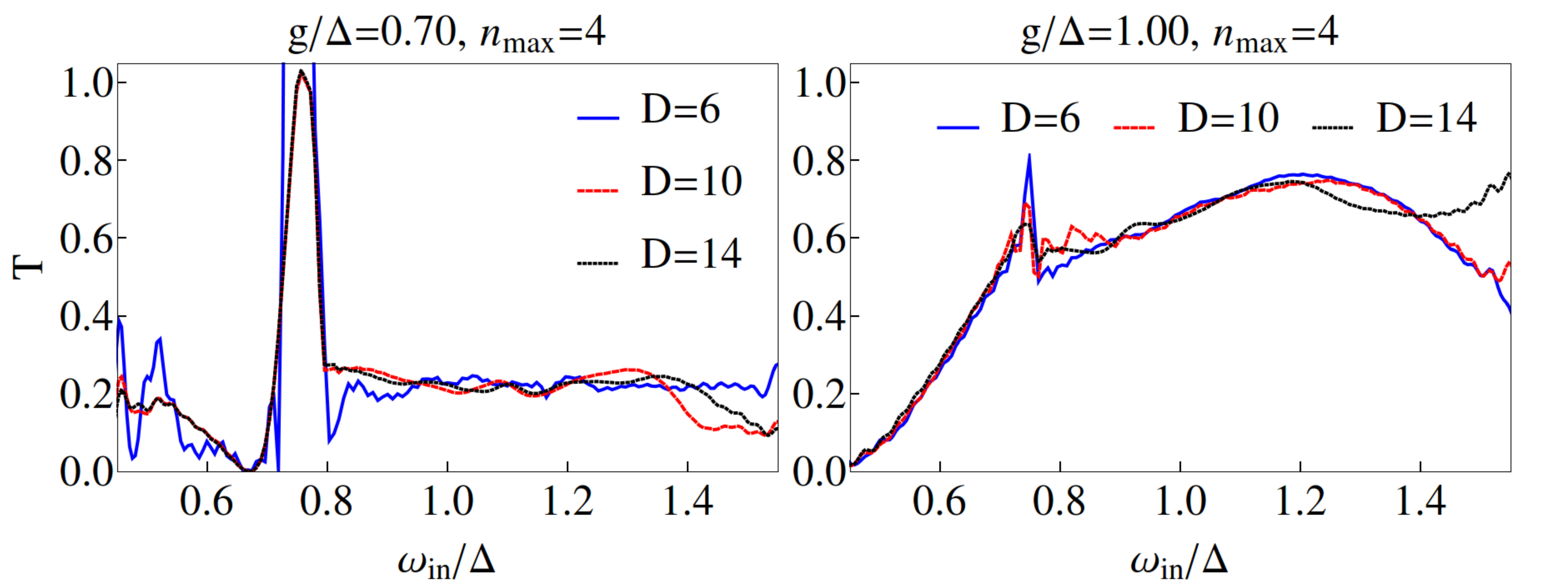}
\caption{{\bf Convergence in $D$.} We check that our results converge in the bond dimension for $D=6$ (blue, solid), $D=10$ (red, dashed) and $D=14$ (black, dotted). For instance, we see that the Fano-like resonance is not described properly if $D$ is not big enough, as we see for $D=6$ in the left pannel. However, for $D=10$ and $D=14$ both results are almost identical.}\label{fig:D}
\end{figure}

In the figure \ref{fig:D} we show the transmission factor vs the incident energy for $n_{max}=4$, for several values of the bond dimension $D=6,10,14$, for $g=0.70$ (left pannel) and $g=1.00$ (right pannel), which is even beyond the aim of this work. As we see, the curves are pretty similar for all $D$ in all the frequency range, except around the Fano-like resonance in the left pannel, where we obtain unphysical results for $D=6$, since $T$ is larger than 1 for those values of $\omega_{in}$.

\begin{figure}[h]
\includegraphics[width=1\columnwidth]{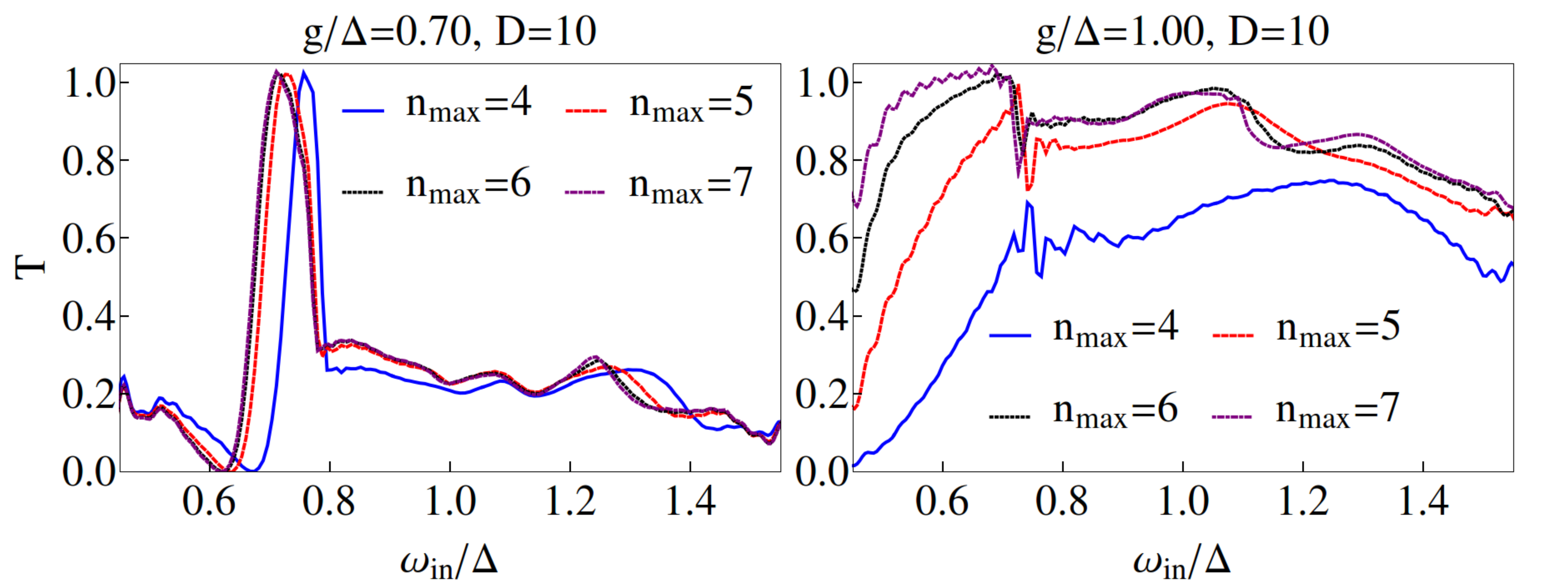}
\caption{{\bf Convergence in $n_{max}$.} Here we do the same as in
  Fig. \ref{fig:D}  with $n_{max}$; $n_{max}=4$ (blue, solid), $n_{max}=5$ (red, dashed), $n_{max}=6$ (black, dotted) and $n_{max}=7$ (purple, dotted-dashed). We find just quantitative differences. For $n_{max}=6,7$, we conclude that the results are highly converged, but the simulations for $n_{max}=4$ capture the phenomenology.}\label{fig:nmax}
\end{figure}

In the figure \ref{fig:nmax} we fix $D=10$ and take $n_{max}=4,5,6,7$, for the same values of the coupling constant $g$. As it is seen, there are not qualitative changes. In the left pannel, for $g=0.70$, we see just a shift in the peak position of the Fano resonance. It is clear the the curve converges for $n_{max}=6$. On the other hand, for $g=1.00$, the only difference is that, as $n_{max}$ increases, the qubit and the electromagnetic field decouple since $T$ goes to $1$ in a really broad region in $\omega_{in}$.

\section{Details of the simulations}

We took chains of $L=480$ cavities, with the qubit interacting with the cavity placed at $j=240$ and the incident wave packet centered at $j=160$, except for the inset of the figure 5, where we placed the qubit interacting with the cavity at $j=460$ and the incident wave packet centered at $380$. The results of the figures 3 and 6 were done with $n_{max}=4$ and $D=10$. The width of the incident wave packet of the figures $4$ and $5$ is $\sigma=2$ (narrow in positions, broad in momenta, to compute the transmission factor for a large range of energies), whereas for the figures $3$ and $6$ we took $\sigma=20$, to see the dynamics of photons with well defined momentum. We took a total time $t=420$ in all the simulations, but in the inset of the figure 5, where we took $t=800$, since that wavepacket between the qubit and the wall goes back and forth again and again.

%\color{blue}
%Se me ocurre que ir\'{i}a bien aqu\'{i} un diagrama de contorno con el factor de transmisi\'{o}n para un determinado $g$ en funci\'{o}n de $\omega_{in}$ y $t$, para ver que converge a una curva bien definida. Me cuesta un momento preparar una gr\'{a}fica as\'{i}.
%
%\color{black}
\section{Frequency shift}

In this section we show that it is possible to describe properly the frequency shift with an approximate calculation. First of all, we consider that the scatterer is the cavity-qubit system and we truncate its Hilbert space just to the ground state and the couple of states which have just one particle in the low coupling regime, that is, the polariton states, which in RWA are

\begin{equation}
|e_{\pm}\rangle=(a_0^\dagger|0\rangle \pm \sigma^+|0\rangle)/\sqrt{2}.\label{polariton}
\end{equation}

Then, a general state in this subspace is

\begin{equation}
|\Psi\rangle = \sum_{n\neq 0}c_n a_n^\dagger |GS\rangle + f_+|\tilde{e}_+\rangle + f_-|\tilde{e}_-\rangle,
\end{equation}
where $\{|\tilde{e}_i\rangle\}$ are the polariton states calculated beyond the RWA for a system comprising just one cavity plus one qubit. Taking the following ansatz we can find the scattering eigenstates

\begin{equation}
c_n=\left\{\begin{array}{l}
e^{ikn} + r_k e^{-ikn}\qquad n<0\\
t_k e^{ikn}\qquad\qquad\quad\,\,\,\, n>0
\end{array}\right.
\end{equation}

Solving the eigenvalue equation $H\ket{\Psi}=E\ket{\Psi}$, we show that the transmission amplitude is

\begin{equation}
t_k=\frac{2iG\sin k}{2e^{ik}G-1},\qquad G:=J\sum_{i=\pm}\frac{|\alpha_{i0}|^2}{\Delta_i-\omega_k}.
\end{equation}

Here, $\Delta_i$ is the gap between $|\tilde{e}_i\rangle$ and $|GS\rangle$ and $\alpha_{i0}=\langle \tilde{e}_i|a^\dagger |GS\rangle$. By imposing $t_k=0$, we find that the resonant energy for perfect reflection is

\begin{equation}
\omega_R=\frac{|\alpha_{+0}|^2 \Delta_- + |\alpha_{-0}|^2 \Delta_+}{|\alpha_{+0}|^2+|\alpha_{-0}|^2}.
\end{equation}

\begin{figure}
\includegraphics[width=0.7\columnwidth]{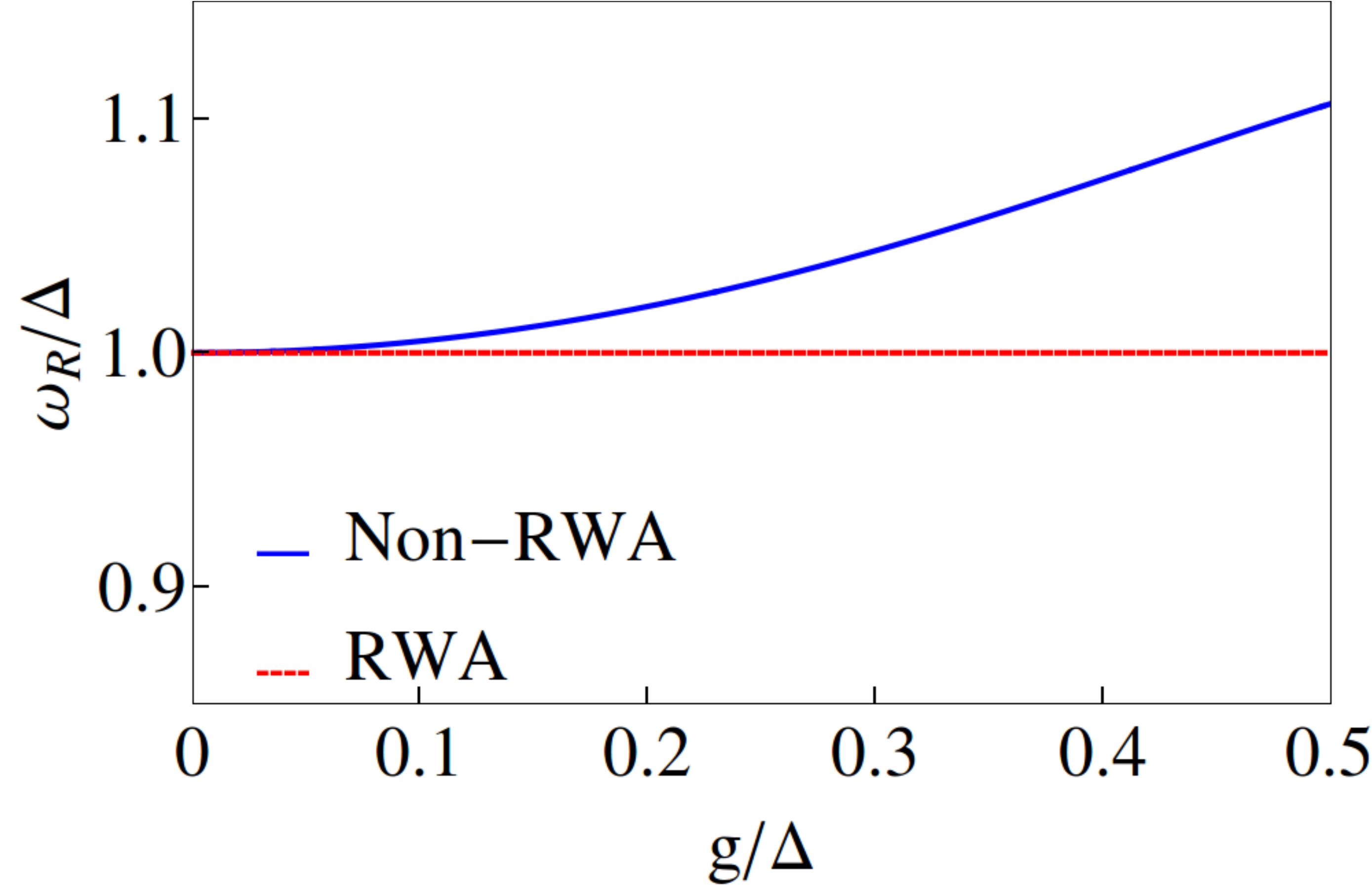}
\caption{{\bf Resonant energy for perfect reflection.} $\omega_R$ shifts to larger values as $g$ increases when computed beyond the RWA (blue, solid line), whereas it remains constant within the RWA (red, dashed line).}\label{fig:omegaR}
\end{figure}

In the RWA, $\alpha_{\pm 0}=1/\sqrt{2}$, and $\Delta_{\pm}=\Delta\pm g$, so $\omega_R=\Delta$. However, counter rotating terms modify both the gaps and the matrix elements, so the resonant frequency shifts, as we plot in the figure \ref{fig:omegaR}. In the manuscript, the same curve is plotted over the figure 4.b, and it fits really well with the numerical result obtained with MPS. A deeper study of this method will be shown elsewhere.

\section{No Raman scattering with linear systems}

\subsection{Linear models:
\\ Definition and first properties}

Let us begin by defining a linear system.
\begin{definition}
A linear model \footnote{We say that this is a linear model because the corresponding Heisenberg equations are linear} consists of a quadratic Hamiltonian of creation and annihilation
bosonic operators ($[a_i, a_j^\dagger] = \delta_{ij}$, $[a_i,a_j]=0$):
\begin{equation}
\label{lin-ham}
H = \sum_{i,j}^{N+s} (\gamma_{i,j} a_{i}^\dagger a_{j}
+
(\beta_{i,j} a_{i} a_{j}
+
{\rm h c})).
\end{equation}
\end{definition}
Here, the matrices $\hat \gamma$  and $\hat \beta$ define the Hamiltonian.
Introducing the vectorial notation:
\begin{equation}
\label{array}
{\bf a}
=
\left (
\begin{array}{c}
a_1
\\
\vdots
\\
a_N

\\
a_{N+1}
\\
\vdots
\\
a_{N+S}
\\
a_1^\dagger
\\
\vdots
\\
a_N^\dagger
\\
a_{N+1}^\dagger
\\
\vdots
\\
a^\dagger_{N+S}
\end{array}
\right ).
\end{equation}
the Hamiltonian (\ref{lin-ham}) can be rewritten as
\begin{equation}
\label{quadratic}
H = {\bf a}^\dagger  \, \hat h \, {\bf a}
,\qquad
\qquad
h=
\left (
\begin{array}{cc}
\hat \gamma  & \hat \beta
\\
\hat \beta ^\dagger & \hat \gamma^{t}
\end{array}
\right ) ,
\end{equation} 
up to an additive constant. Notice that imposing (\ref{lin-ham}) and its equivalent
(\ref{quadratic})  to be Hermitian, we need to fulfill: $\gamma = \gamma^\dagger$. On the other hand, we can take $\beta$ symmetric without loss of generality.
\subsubsection{Normal modes}

Diagonalized within Bogolioubov-Valatin transformation \footnote{We
  do not show that a BV transformation diagonalizes any general
  quadratic form.  Indeed, we do not know.  Typical hamiltonians: {\it
    e.g. }$\gamma$ and $\beta$ real  do.
  }, the
Hamiltonian can be written (up to an irrelevant constant) as:
\begin{equation}
\label{H-normal}
H = \sum \epsilon_l \alpha^\dagger _l \alpha_l ,
\end{equation}
with $[\alpha_l, \alpha_m^\dagger]=1$ and $\epsilon_l > 0$ $\forall l$ if the modell is well behaved.

The operators $\alpha_l^\dagger$, which can be understood as generators of generalized normal modes,
provide a simple representation of the ground state:
\begin{equation}
\label{alpha-GS}\alpha_l | G S \rangle = 0.
\end{equation}

It is important to notice that the $\alpha^\prime$s and the $a^\prime$s,
are linearly related:
\begin{equation}
\label{alpha-A}
\alpha_l := \sum_j (\chi_{lj} a_j + \eta_{lj} a_j^\dagger).
\end{equation}

This normal mode representation provides a natural and convenient way
of labeling the states {\it \`{a} la Fock}.  The eigenstates of $H$
can be written as:
\begin{equation}
| n_1, ..., n_L \rangle \sim (\alpha_1^\dagger)^{n_1} \, \hdots
(\alpha_1^\dagger)^{n_L} | 0, \hdots, 0\rangle
\; 
\end{equation}
where 
\begin{equation}
\label{GS-normal}
|GS \rangle =  | 0, \hdots, 0\rangle
\end{equation}
These states are mutually orthogonal:
\begin{equation}
\langle n_1, ..., n_L | m_1, ..., m_L \rangle = \prod_i^L \delta_{n_i, m_i}.
\end{equation}

{\bf Remark 1}  Notice that Hamiltonian (\ref{H-normal}) splits
the Hilbert space in orthogonal sectors where the total 
\begin{equation}
N_\alpha =
\sum \alpha_l^\dagger \alpha_l 
\end{equation}
is fixed. 
This is true despite the fact that the original Hamiltonian ($\hat \beta \neq
0)$ is not {\it number conserving}  in the $a'$s: $[H,\sum_j a^\dagger_j a_j ] \neq 0$.
Then, the number of excitations $N_\alpha$ turns to be a good quantum number. We will refer to it as $\alpha$-particles or $\alpha$-excitations.

{\bf Remark 2}  A quadratic form in the bosonic fields, as
(\ref{quadratic}) conserves the parity: $P = {\rm e}^{i \pi \sum
  a_j^\dagger a_j}$, $[H, P]= 0$.  Trivially, the Hamiltonian
also conseves the parity in the $\alpha '$s.

\subsection{Scattering input and time evolution}

Let us consider a single-photon input state:
\begin{equation}
\label{in}
| \psi_{in} \rangle
= 
\sum_j \phi_j a_j^\dagger | GS \rangle
= 
\sum_l \tilde{\phi}_l \alpha_l^\dagger  | 0, \hdots , 0 \rangle.
\end{equation}

The second equality holds since $a_j^\dagger$ depends linearly on $\alpha_l$ and $\alpha_l^\dagger$ and $\alpha_l$ annihilates the ground state (\ref{alpha-GS}). This is a key point: The initial state is a well
defined single particle state (in $\alpha$-particles).

Then, as the number of $\alpha$-excitations is a conserved quantity, the time evolution is restricted to the {\it one $\alpha$-excitation or $\alpha$-particle
  level}:
\begin{equation}
\label{N-conserved}
N_{\alpha} |\psi_{in} \rangle = |\psi_{in} \rangle \Rightarrow N_\alpha {\rm e}^{-i H t} | \psi_{in} \rangle = {\rm e}^{-i H t} | \psi_{in} \rangle .
\end{equation}
%\begin{align}
%\nonumber
%{\rm e}^{ - i H t}
%| \phi_{in} \rangle
%&=
%\sum \psi_j {\rm e}^{-i H t} 
%\, a_j^\dagger \,  {\rm e}^{+i H t} \,  {\rm e}^{-i
 % H t} | G S \rangle
%\\ \nonumber
%&=
%\phi_j  \, a_j^\dagger (-t) \,  | GS \rangle
%\\ 
%&= 
%{\bf \Phi}^ t (t) \, {\bf A} | GS \rangle
%\end{align}
%here $a_j^\dagger (-t)$ is the usual Heisenberg picture evolution.
%In the last equality we have made use of the linearity and introduce
%some notation: $\bf Phi (t)^t = (\phi_1 (t)$

\subsection{No Raman scattering. A theorem}

\begin{theorem}
Given the single particle input state (\ref{in}) there is not
Raman scattering in linear optics. In other words the output frequency equals the
input one.
\end{theorem}

\begin{proof}
By {\it Reductio ad absurdum}:\newline

Let us write the input state (\ref{in}) in momentum space:
\begin{equation}
| \psi_{in} \rangle =
\sum_k \phi (k - k_0) a_k^\dagger | GS \rangle,
\end{equation}
where $\phi(k-k_0)$ is a wave packet whose momentum is well defined
around $k_0$. The output state is a combination of transmitted and
reflected states. If there is Raman scattering:
\begin{align}
| \psi_{out} \rangle &  = | \psi_{out}^t \rangle + | \psi_{out}^r \rangle,\nonumber \\
\label{out-t}| \psi_{out}^t \rangle & = \sum_k t_k \phi(k - k_0)a_k^\dagger |GS\rangle + \sum_k \phi^\prime (k - k_1)a_k^\dagger |EXC\rangle,\\
\label{out-r}| \psi_{out}^r \rangle & = \sum_k r_k \phi(k + k_0)a_k^\dagger |GS\rangle + \sum_k \phi^\prime (k + k_1)a_k^\dagger |EXC\rangle,
\end{align}
where $|EXC\rangle$ is an excited state and $k_1$ is the new momentum. Energy
conservation forces that:
\begin{equation}
\label{energy}\omega_{k_0} + E_{GS} =  \omega_{k_1} + E_{EXC}.
\end{equation}

The excited state $|EXC\rangle$ must have the same parity as $|GS \rangle $ (see {\bf Remark 1}). In addition, it must be eigenstate of $N_\alpha$. Then $N_{\alpha} | EXC \rangle = 2 n| EXC \rangle$ $(n \geq
1)$. Rewriting the second term of (\ref{out-t}) (equivalently for
(\ref{out-r})) in terms of the $\alpha$ operators \footnote{Since
  $\alpha_l$ and $\alpha_l^\dagger$ are linearly related to $a_j$ and
  $a_j^\dagger$ and the relation is invertible and, since $a_j$ are
  linear in $a_k$, $a_k$ and $a_k^\dagger$ are linear in $\alpha_l$
  and $\alpha_l^\dagger$; . $\theta$ and $\mu$ are the matrices which
  relate $a_k$ to $\alpha_l$ and $\alpha_l^\dagger$.}:
\begin{equation}
\label{freq-shift}\sum_k \phi^\prime( k- k_1)\sum_l (\theta_{kl}\alpha_l + \mu_{kl}\alpha_l^\dagger)|EXC\rangle.
\end{equation}

Trivially the second term in (\ref{freq-shift}) does not belong to the one $\alpha$-excitation
sector. Which is a contradiction, since (\ref{H-normal}) does not
couple different  $\alpha$-sectors. On the other hand, the first term in (\ref{freq-shift}) belongs to the one $\alpha$-particle sector, so it can be written as a wave packet created over the ground state. Then, because of energy conservation (\ref{energy}), that wave packet has momentum $k_0$.
This ends the proof.
\end{proof}

\bibliographystyle{apsrev4-1}
\bibliography{scattering_eduardo}

\end{document}